\documentclass{article}
\usepackage{graphicx}
\usepackage{epsfig}
\usepackage{amsfonts}
\newcommand{\beq}{\begin{equation}}
\newcommand{\eeq}{\end{equation}}
\newcommand{\beqa}{\begin{eqnarray}}
\newcommand{\eeqa}{\end{eqnarray}}
\newcommand{\beqar}{\begin{eqnarray*}}
\newcommand{\eeqar}{\end{eqnarray*}}
 
\begin{document}
\title{ Nonabelian solutions in a Melvin magnetic universe} 

\author{{\large Burkhard Kleihaus$^1$,}  
{\large Jutta Kunz$^2$} %
 and  {\large Eugen Radu$^2$}  
 \\
 \\
{\small $^1$ ZARM, Universit\"at Bremen, Am Fallturm,
D--28359 Bremen, Germany}
 \\
{\small $^2$ Institut f\"ur Physik, Universit\"at Oldenburg, Postfach 2503
D-26111 Oldenburg, Germany} }

\maketitle

\begin{abstract}
We show the existence
of $D=4$   nonabelian solutions
approaching asymptotically a dilatonic Melvin spacetime background.
An exact solution generalizing the Chamseddine-Volkov soliton
for a nonzero external U(1) magnetic field is also reported.
\end{abstract}

\section{Introduction}
According to the so called "no-hair" conjecture, an asymptotically flat,
stationary black hole is uniquely
described in terms of a small set of asymptotically measurable quantities. 
However, in recent years counterexamples to this conjecture 
were found in several theories, most of them containing  
nonabelian matter fields. 
The first nonabelian "hairy" black hole solutions 
within the framework of  SU(2)  
Einstein-Yang-Mills (EYM) theory, were presented in \cite{89}.
Although the new solutions were static and have vanishing Yang-Mills (YM) 
charges, they were different from the Schwarzschild black hole and,
therefore, not characterized by their total mass. 
Remarkably, 
in the limit of zero event horizon radius of these configurations, 
the globally regular, particle-like solutions originally found
in \cite{Bartnik:1988am} are recovered.

It is however worth inquiring what happens with these solutions
if we drop the assumption of asymptotic flatness.
The asymptotically (anti)-de Sitter solutions which are found 
for a nonzero cosmological constant enjoyed considerable 
attention in the last years and present many interesting features \cite{Winstanley:1999sn}.

Another interesting example of non-asymptotically flat solutions in general relativity
is given by the Melvin magnetic universe,
describing a bundle of magnetic flux lines in
gravitational-magnetostatic equilibrium \cite{Melvin:1963qx}.
This solution has a number of interesting features,
providing the closest approximation in general relativity for an uniform magnetic field.
The nonsingular  nature of this solution (at the cost of losing the asymptotic
flatness) motivated Melvin to refer to his solution as a magnetic "geon".

There exists a fairly extensive literature on the properties of
this magnetic universe, starting with a study by Thorne, which investigates also
its dynamical behaviour under arbitrary large radial perturbations \cite{thorne}.
Various generalizations of this type of solution  have been proposed
(see \cite{KaBu} for a review and relevant references), 
particularly interesting being the Melvin solution in Kaluza-Klein (KK) theory.
This configuration derives from a flat five dimensional 
spacetime by performing a U(1) reduction 
with a twist in the identifications \cite{Dowker:1994up,Dowker:1995gb},
the four dimensional theory containing an extra dilaton field. 
An exact solution of Einstein-Maxwell equations describing a black hole
in a background Melvin universe was constructed  by Ernst \cite{Ernst1}, and admits also a straightforward
generalization in the  KK case \cite{Dowker:1994up}.

It is therefore natural to ask whether 
the well-known hairy
black hole solutions admit  generalizations with Melvin-type asymptotics
and what new effects emerge due to the presence of a background magnetic field.
The main purpose of this paper is to present such solutions 
in EYM-Higgs-U(1)-dilaton theory,
which approach asymptotically a Melvin background.

%
\section{General framework }
We consider the following action in four spacetime dimensions 
\begin{eqnarray}
\label{action4}
I_4=\frac{1}{4\pi}\int d^{4}x\sqrt{-\gamma }\Big[ 
\frac{\mathcal{R} }{4}
-\frac{1}{2}\nabla_{i}\psi \nabla^{i}\psi 
-e^{2a\psi}\frac{1}{4}f_{ij }f^{ij } 
-e^{2a\psi/3 }\frac{1}{4} 
 {\cal F}^{'I}_{ij}  {\cal F} ^{'I ij } 
-e^{-4a\psi/3 }\frac{1}{4} D_i\Phi^I D^i\Phi^I
\Big],
\end{eqnarray}
which describes a gravitating system with  
a scalar triplet $\Phi^I$ ($I=1,2,3$), an
SU(2) Yang-Mills (YM) potential ${\cal A}_{i}^I$ 
(with field strength ${\cal F}_{ij}^{I}=
\partial_i {\cal A}_j^I-\partial_j {\cal A}_i^I
+\epsilon^{IJK}{\cal A}_i^J {\cal A}_j^K$),  
an abelian potential
${\cal W}_{i}$ 
($f_{ij}=\partial_i {\cal W}_j -\partial_j {\cal W}_i$ being the corresponding field strength)
and a dilaton field $\psi$, 
 $a$ being the dilaton coupling constant, and
we note  
${\cal F}_{ij}^{'I}={\cal F}_{ij}^{I}+2\Phi^{I} f_{ij}$.
This expression of the action has  a  higher dimensional origin and is motivated 
in the next Section.

\subsection{The dilaton Melvin solution}
The Melvin solution
in Einstein-Maxwell-dilaton theory is found 
for vanishing SU(2) and triplet scalar fields, 
${\cal F}_{ij}=\Phi^I=0$, and reads \cite{Dowker:1994up}
\begin{eqnarray}
\label{Melvin}
ds^2=\Lambda^{\frac{2}{1+a^2}}(dr^2+r^2 d\theta^2 - dt^2)
+\Lambda ^{-\frac{2}{1+a^2}}r^2 \sin^2 \theta d \varphi^2,~~
{\rm with}~~\Lambda =1+(\frac{1+a^2}{4})B_0^2 r^2 \sin^2 \theta,
\end{eqnarray}
the dilaton and the $U(1)$ potential being
\begin{eqnarray}
\label{alte}
~~e^{2a(\psi-\psi_0)}=\Lambda^{\frac{2a^2}{1+a^2}}
,~~{\cal W}_{i}dx^i=\frac{e^{-a\psi_0} B_0r^2 \sin^2 \theta}{2\Lambda }d\varphi.
\end{eqnarray}
The solution  is parametrized by $\psi_0$, 
the value of the scalar field on the symmetry axis and
$B_0$, which characterizes the central strength of the magnetic field.
Although not asymptotically flat, the geometry of this solution is singularity
free and geodesically complete.
A curious property of (\ref{Melvin})-(\ref{alte}) is that the total flux
\begin{eqnarray}
\label{flux0}
 \Phi_m=\oint_{\infty} {\cal W}_{\varphi}=e^{-a\psi_0} \frac{4\pi}{1+a^2}\frac{1}{B_0}
\end{eqnarray}
is finite and inversely proportional to $B_0$.
(The total magnetic flux for this cylindrically symmetric solution
 is obtained by integrating over the entire
physical area perpendicular to the $z-$axis, with $z=r\cos \theta$ \cite{Melvin:1963qx}).
However, in the limit $B_0 \to 0$, even if the geometry becomes flat
and the field strength goes to zero at the centre, the total flux diverges.

The  solution describing a Schwarzschild black hole immersed in the 
dilatonic Melvin universe is a straightforward
generalization of (\ref{Melvin}), (\ref{alte}) and has a line element  \cite{Dowker:1994up}
\begin{eqnarray}
\label{Ernst}
ds^2=\Lambda ^{\frac{2}{1+a^2}}\bigg(\frac{dr^2}{1-\frac{2M}{r}}+
r^2 d\theta^2 -(1-\frac{2M}{r}) dt^2\bigg)
+\Lambda ^{-\frac{2}{1+a^2}}r^2 \sin^2 \theta d \varphi^2,
\end{eqnarray}
with the same expressions for $\Lambda$, dilaton and U(1) potential 
(the $\psi=0$ case was first discussed in \cite{Ernst1}).

The constant $M$ which enters the line element (\ref{Ernst}) corresponds to the black hole's
mass.
This axially symmetric solution contains an event horizon
at $r=2M$ as in the Schwarzschild vacuum case, but is not asymptotically
flat owing to the gravitational effects of the magnetic field.
It is evident that standard Kruskal coordinates may be introduced 
in order to extend the solution across the event horizon,
the only singularity occurring at $r=0$.
More details on this solution can be found e.g. in \cite{KaBu},\cite{Hiscock:1980zf},\cite{Radu:2002pn}.

\subsection{Nonabelian Einstein-Yang-Mills-dilaton solutions}
For a vanishing U(1) and  triplet scalar fields, ${\cal W}_i=\Phi^I=0$, one finds 
a different class of  solutions,
corresponding to dilaton generalizations of the SU(2)-EYM hairy black holes 
 \cite{89}.
In the simplest, spherically symmetric case, these configurations 
are usually described by using a line element 
\begin{equation} 
\label{BK}
ds^2=
\frac{dr^2}{N(r)}+r^2(d\theta^2+\sin^2 \theta d\varphi^2)-\sigma^2(r)N(r)dt^2,~~~
{\rm with}~~ N(r)=1-\frac{2m(r)}{r},
\end{equation}
$m(r)$ corresponding
to the total  mass-energy within the radius $r$,
and a SU(2) nonabelian potential
\begin{equation} 
\label{YMans}
{\cal A}_i dx^i=w(r)(\tau_1 d\theta+\tau_2 \sin \theta d \varphi)+\tau_3\cos \theta d \varphi,
\end{equation}
$\tau_i$ being the Pauli matrices.
The metric functions $m(r)$, $\sigma(r)$, the gauge
potential function $w(r)$ and the dilaton function $\psi(r)$ are solutions of the equations
\begin{eqnarray}
\label{eqs}
m' = 2 \Big(e^{2a \psi/3}
 (w'^2 N+\frac{(w^2-1)^2}{2r^2})
+\frac{r^2}{2}N\psi'^2  \Big),
~~~~
\sigma'=\frac{2}{r}\Big(e^{2a \psi/3} w'^2
+\frac{1}{2}\psi'^2r^2 \Big),
\\
\nonumber
\left(\sigma e^{2a \psi/3}Nw' \right)'=
\sigma e^{2a \psi/3} \frac{w(w^2-1)}{r^2},
~~~
\left(Nr^2 \sigma \psi'\right)'=\frac{2a}{3} \sigma  e^{2a \psi/3}  
\left(
w'^2 N+\frac{(w^2-1)^2}{2r^2}\right),
\end{eqnarray}
(where a prime denotes a derivative with respect to the radial coordinate $r$),
with suitable boundary conditions.

Although no exact nonabelian solutions of the above equations
are known,  Refs. \cite{Lavrelashvili:1992ia, Donets:1992zb}
present both analytical and numerical arguments for the existence
 of a discrete family of black hole solutions uniquely characterized
 by the number of nodes $p$ of the function $w(r)$, with $p \geq 1$.
Nontrivial solutions are found for any value of the dilaton coupling constant $a$,
the dilaton field vanishing asymptotically.
 
 These  solutions approach asymptotically the Minkowski spacetime background ($m(r)\to M,~\sigma(r) \to 1$)  and 
have no global nonabelian charge (although their dilaton charge is nonzero). 
The black hole configurations exist for any value of the event horizon radius $r_h$.
 The gauge potential $w$ interpolates between   $w(r_h)=w_0$ (with $|w_0|<1$) 
 and $w(r\to \infty)=\pm 1$,
the Schwarzschild  solution being recovered for $w(r)=\pm 1$ (a pure gauge field), $\sigma=1$,
$\psi=0$ and $m(r)=M$.
In the limit $r_h \to 0$, a dilatonic generalization of the 
Bartnik-McKinnon EYM solutions  \cite{Bartnik:1988am} is approached.

The thermodynamics of the EYM-dilaton black holes can be discussed in the standard way
(see e.g. \cite{Visser:1993qa}); it turns out that
their entropy is one quarter of the event horizon area $S=\pi r_h^2$, while their Hawking temperature  
is $T_H=\sigma(r_h) N'(r_h)/(4\pi)$.

%
\section{The twisting procedure and new solutions}
The purpose of this Section is to present a family of solutions which
extremizes the action  (\ref{action4}),
keeping the basic features of both the Melvin universe (\ref{Melvin}) and the nonabelian
solutions (\ref{BK}), (\ref{YMans}).

Here we restrict to the case of a dilaton coupling constant $a=\sqrt{3}$,
in which case the 
nonabelian solutions (\ref{BK}) can be uplifted to become  
solutions of the SU(2) EYM equation in five dimensions \cite{Volkov:2001tb,Brihaye:2005pz},
extremizing the action
\begin{equation}
\label{action5}
I_5=\frac{1}{4\pi}\int d^{5}x\sqrt{-g }\Big(\frac{1}{4}R
-\frac{1}{4 } F_{\mu \nu }^IF^{I\mu \nu} \Big).
\end{equation}
In a five-dimensional perspective, the solutions of the $D=4$ EYMd equations (\ref{eqs}) with $a=\sqrt{3}$
describe hairy black strings
 or  nonabelian vortices, with a line element
($x^5$ being the extra-direction which is supposed to be compact and with a unit length) 
\begin{equation} 
\label{BK-5D}
ds_5^2 = e^{- a\psi } \left(\frac{dr^2}{N }+r^2(d\theta^2+\sin^2 \theta d\varphi)
-\sigma^2 N dt^2\right)+ e^{ 2a\psi }(dx^5)^2 ,
\end{equation}
and the same SU(2) ansatz (\ref{YMans}),  i.e.  a vanishing fifth component of the nonabelian
potential, $A_5=0$.

The way to introduce a $D=4$ magnetic field in a KK setup 
involves twisting the compactification direction.
Following \cite{Dowker:1994up,Dowker:1995gb} one shifts the coordinate
$\varphi \to \varphi+B_0 x^5$ (with $B_0$ an arbitrary real constant), 
and reidentifies points appropriately.
The next step is to consider the KK reduction with respect to the Kiling vector
$\partial/\partial x^5$, according to the generic prescription
\begin{equation} 
\label{KK5D}
ds_5^2 = e^{- a\psi }\gamma_{ij}dx^{i}dx^{j}
 + e^{ 2a\psi }(dx^5 + 2{\cal W}_{i}dx^{i})^2,
\end{equation}
$\gamma_{ij}dx^{i}dx^{j}$ being the four dimensional line element and ${\cal W}_{i}$ the U(1) potential.
For the  reduction of the YM action term,
a convenient $D=5$ SU(2) ansatz is
\begin{eqnarray}
\label{SU2}
A_{\mu}^Idx^{\mu}={\cal A}_{i}^Idx^{i}+\Phi^I (dx^5+2 {\cal W}_i dx^i),
\end{eqnarray}
where  
${\cal A}_{i}^I$ is a purely four-dimensional YM gauge field potential,
while  $\Phi^I$ corresponds after the dimensional reduction to a
triplet Higgs field.
It can be verified that the KK reduction of the action (\ref{action5}) with respect 
to the $x^5-$direction, taken according to (\ref{KK5D}), 
(\ref{SU2}), yields the four-dimensional
action (\ref{action4}).

Therefore, upon reduction, the new 
$D=4$ solutions based on the configurations in Section 2.2, have a line element
\begin{equation} 
\label{solution}
ds^2=\sqrt{\Lambda }\left(\frac{dr^2}{N }+r^2 d \theta^2-\sigma^2 N dt^2\right)
 +\frac{r^2 \sin^2 \theta}{\sqrt{\Lambda }} d\varphi^2,
 ~~{\rm with~~}\Lambda= 1+e^{-3 a\psi} B_0^2r^2 \sin^2 \theta,
\end{equation}
the only nonvanishing component of the U(1) potential vector ${\cal W}_i$ being 
\begin{eqnarray}
\label{W}
{\cal W}_{\varphi}=\frac{e^{-3 a\psi} B_0r^2 \sin^2 \theta}{2\Lambda }.
\end{eqnarray} 
The new $D=4$ dilaton field $\bar \psi$ is
\begin{eqnarray}
\label{dil-d4}
 \bar \psi =  \psi+\frac{1}{2a}\log \Lambda,
\end{eqnarray} 
while the four dimensional YM field is given by 
\begin{eqnarray}
\label{A4}
{\cal A}_{i}dx^{i}=w(\tau_1 d\theta+\tau_2 \sin \theta d \varphi)+\tau_3 \cos \theta
d \varphi
 -2B_0 (\tau_2w \sin \theta +\tau_3 \cos \theta){\cal W}_{\varphi}d \varphi~.
\end{eqnarray} 
Different from the seed solutions, the new configurations have a nonvanishing Higgs field
\begin{eqnarray}
\label{higgs}
\Phi=B_0(w \sin \theta \tau_2+\cos \theta \tau_3).
\end{eqnarray} 
It can easily be seen that for a vanishing nonabelian
matter content $(w=\pm 1)$, the dilatonic Ernst solution \cite{Ernst1} 
describing a Schwarzschild black hole in a
Melvin background is recovered, while setting $B_0=0$ 
leads us back to the EYMd seed solution (\ref{BK}), (\ref{YMans}).

A different type of configuration is found for $w(r)=0$,
describing a magnetic monopole black hole placed in a Melvin universe
(note that here the four dimensional geometry has a closed form expression \cite{Dowker:1994up}, for a different
parametrization instead of (\ref{BK}), however).

For the generic case,
one can  see that the causal structure of the seed EYMd solution is not changed by the
twisting procedure. Supposing one starts with an initial EYMd  hairy black hole solution, 
one finds
that the Melvin-type metric (\ref{solution}) describes,
in terms of the usual definitions, a black hole,
with an event horizon and trapped surfaces.
It has a horizon located at $r=r_h$ (where $N(r_h)=0$), which is independent 
of the value of the magnetic field strength. 
A globally regular configuration (which differs from the Melvin solution) is found in the limit of zero event horizon radius.
For $r \to \infty$, the line element (\ref{solution}) 
approaches the Melvin background (\ref{Melvin}) with $a=\sqrt{3}$.

 Similar to the initial YM ansatz  (\ref{YMans}), the YMH fields (\ref{A4}), (\ref{higgs})
 are written in a singular gauge.
A regular form is obtained after applying a 
 gauge transformation
$S=e^{i\pi\tau_3/4}e^{i\theta\tau_2/2}e^{i\varphi \tau_3/2}$.
Their new expression, written in terms of a general ansatz  
used before in the literature
on axially symmetric nonabelian solutions (see e.g. \cite{Kleihaus:1997ws,Hartmann:2001ic})
is
\begin{eqnarray}
\label{ansatz-YMn}
 A_\mu dx^\mu
=
\left[\frac{H_1}{r}dr
 +(1-H_2)d\theta \right] {\tau_\varphi } 
-\sin\theta\left[H_3  {\tau_r } 
            +(1-H_4)  {\tau_\theta } \right]   d\varphi~,~~
	    \Phi= (\phi^r\tau_r+\phi^\theta\tau_\theta)~,
\end{eqnarray}
where $H_1=0$, $H_2=w$, $H_3=-2B_0{\cal W}_{\varphi}\cot \theta$, 
$H_4=w(1-2{\cal W}_{\varphi}B_0)$ and 
$\phi^r=B_0\cos \theta $, $\phi^\theta=-w B_0 \sin \theta$.
As usual, the symbols $\tau_r$, $\tau_\theta$ and $\tau_\phi$ in the above relation
denote the dot products of the cartesian vector
of Pauli matrices, $\vec \tau = ( \tau_1, \tau_2, \tau_3) $,
with the spatial unit vectors $
\vec e_r =
(\sin \theta \cos  \phi, \sin \theta \sin  \phi, \cos \theta), 
\vec e_\theta  = 
(\cos \theta \cos  \phi, \cos \theta \sin  \phi,-\sin \theta),   
\vec e_\phi = (-\sin \phi, \cos \phi,0), $
respectively. 

The matter fields of the new solution
possess a nontrivial dependence on the polar coordinate $\theta$.
The modulus of the Higgs field $|\Phi|=\sqrt{\Phi^I\Phi^I}$
approaches a constant value at infinity and vanishes on $p$ circles in the $xy$-plane ($\theta=\pi/2$),
which are located at the zeros of the seed gauge potential $w(r)$.
(The occurrance of asymptotically flat vortex ring solutions in a pure EYMH theory has
 been  noticed in \cite{Kleihaus:2003xz} for a set of monopole-antimonopole solutions).
Given the non asymptotically flat character of the spacetime, the interpretation of 
the matter field configurations in this solution is not obvious.
However, since the modulus of the Higgs field 
is constant at infinity, as in the asymptotically flat case,
we suggest that the 't Hooft electromagnetic field strength tensor
 \begin{equation}
 \label{emtH}
\mathsf{F}_{\mu\nu} =
\varepsilon_{IJK}\hat{\Phi}^I 
 \partial_\mu \hat{\Phi}^J \partial_\nu \hat{\Phi}^K 
+ \partial_\mu(\hat{\Phi}^IA_\nu^I)
-\partial_\nu(\hat{\Phi}^IA_\mu^I) \  ,
\end{equation}  
(where $\hat{\Phi}^I$ is the normalized Higgs field)
might be used to analyze the solutions. 
Then, following \cite{Kleihaus:1999sx}, 
one would evaluate the total nonabelian
magnetic charge of the configurations, by integrating the 
't Hooft electromagnetic field strength tensor,
\begin{equation}
\mathsf{F}_{\theta\varphi}= 
( (1-2B_0{\cal W}_\varphi)
\sqrt{w^2 \sin^2 \theta+ \cos^2 \theta}~)_{,\theta}
\ .
\label{Hooft}
\end{equation}
Thus the magnetic charge inside a 
closed surface ${\cal S}$ would be expressed as  
$m = \frac{1}{V({\cal{S})}} 
\int_{{\cal S}}{ \mathsf{F}_{\mu\nu} }dx^\mu dx^\nu ,$
which turns out to vanish for the new solution (although locally the magnetic
charge density would be nonzero). 

To interpret the new solution we now
suggest to consider the asymptotic expansion of 
the function $w(r)=\pm \left( 1 -\frac{c}{r} \dots \right)$
in the 't Hooft field strength tensor (\ref{Hooft}),
yielding
$\mathsf{F}_{\theta\varphi}=((1-2B_0{\cal W}_\varphi) 
(1 - \frac{c \sin^2 \theta}{r} + O(\frac{1}{r^2})~))_{,\theta}$
and compare with the gauge potential of a magnetic dipole
with dipole moment $\mu$,
$\tilde {\cal W}_\varphi =  \frac{\mu \sin^2 \theta}{r} $ 
\cite{Kleihaus:2003xz,Kleihaus:1999sx}.
The analogous functional dependence then hints at the
possibility to interpret the new solution as a magnetic dipole
with dipole moment $\mu=-c$, immersed in a Melvin background.
In the asymptotically flat case discussed in \cite{Kleihaus:2003xz},  
the vortex ring solutions (where the Higgs field
vanishes on one or more rings) analogously correspond to magnetic dipoles.

A computation of the thermodynamic properties of the  solution (\ref{solution})-(\ref{higgs})
 can  be done by applying the same approach
as for the $B_0=0$ case.
The computation of the mass and total Euclidean action is done with respect to the Melvin 
background (\ref{Melvin}) (with $a=\sqrt{3}$).
The instanton that enters the calculation of the gravitational action is obtained by setting
$\tau=it$ in (\ref{solution}).
Similar to the pure Einstein-Maxwell-dilaton case \cite{Radu:2002pn},
it follows that the thermodynamic properties of these black holes
are not affected by the background U(1) magnetic field.
In particular
we find the same entropy and mass as for the asymptotically flat configurations;
the value of the Hawking temperature is also unchanged. 
A similar behaviour has been noticed in \cite{Radu:2002pn} 
for the Ernst solution (\ref{Ernst}).
Therefore, this seems to be a generic property of static black hole
solutions in a background $U(1)$ magnetic field extending to infinity.

%
\section{Chamseddine-Volkov soliton in a background magnetic field}

The procedure above may be applied to other nonabelian solutions
with a higher dimensional origin.
A particularly interesting case is given by the Chamseddine-Volkov
solution \cite{Chamseddine:1997nm,Chamseddine:1998mc} 
of the ${\cal N}=4,~D=4$ Freedman-Schwarz gauged supergravity model \cite{Freedman:1978ra}.
This exact solution is globally regular, 
preserves $1/4$ of the initial supersymmetry of
the Freedman-Schwarz model and has unit magnetic charge.
Its ten-dimensional lift was shown 
to represent 5-branes wrapped on a shrinking $S^2$ \cite{Chamseddine:1998mc}.
As conjectured by Maldacena and Nu\~nez, this solution provides 
a holographic description for ${\cal N}=1,~D=4$ super-Yang-Mills theory \cite{Maldacena:2001yy}. 
 
The four dimensional Chamseddine-Volkov solution in \cite{Chamseddine:1997nm,Chamseddine:1998mc}  can be uplifted to
$D=5$ \cite{Chamseddine:2001hk}
to become a solution of a consistent truncation of the ${\cal N}=4$ Romans' model 
\cite{Romans:1985ps}
with an action 
\begin{eqnarray}
\label{action52} 
I_5=\frac{1}{4 \pi}\int  d^5x   \sqrt{-g}  \Big
(\frac{1}{4} R
-\frac{1}{2}\partial_\mu\phi \,\partial^\mu\phi -\frac{1}{4}{\rm e}^{2\sqrt{2/3}\phi}
F^{I}_{\mu\nu} F^{I \mu\nu}
+\frac{1}{8}e^{-2\sqrt{2/3}\phi}\Big) ,
\end{eqnarray}
 with $F^{I}_{\mu\nu}$ the SU(2)  YM  field strength.
The uplifted  Chamseddine-Volkov solution reads \cite{Chamseddine:2001hk}
\begin{eqnarray}
\label{CV-5D}
ds^2=r_0^2e^{2\nu}\left(-dt^2+dr^2+Y (d\theta^2+\sin^2 \theta d \varphi^2)+(dx^5)^2\right),
{\rm with~}
Y=2r\coth r -\frac{r^2}{\sinh^2 r}-1, 
~
e^{6\nu}=\frac{\sinh^2 r}{Y}, 
\end{eqnarray}
$r_0$ being an integration constant, a dilaton field 
\begin{eqnarray}
\phi=\phi_0+\sqrt{\frac{3}{2}}\nu
\end{eqnarray}
and a SU(2) field given by (\ref{YMans}), with $w= {r}/{\sinh r}.$ 
This configuration
is neither asymptotically AdS nor asymptotically flat,
a common situation in the presence of a Liouville dilaton potential \cite{Chan:1995fr,Cai:1997ii}.

To generate a nontrivial $D=4$ Melvin-type solution, one twists again the five dimensional
configuration $\varphi \to \varphi+B_0x^5$, and considers the KK reduction along the $x^5-$direction.
Thus we find that (\ref{action52})
leads to the four dimensional action, which different from (\ref{action4}),
contains two dilatons with a nontrivial potential
\begin{eqnarray}
\label{new-CV-4D-action}
I_4=\frac{1}{4\pi}\int d^{4}x\sqrt{-\gamma }\Big[ 
\frac{\mathcal{R} }{4}
-\frac{1}{2}\nabla_{i}\psi \nabla^{i}\psi
-\frac{1}{2}\nabla_{i}\phi \nabla^{i}\phi
-e^{2\sqrt{3}\psi}\frac{1}{4}f_{ij }f^{ij } 
-e^{2\psi/\sqrt{3}+2\sqrt{2/3} \phi}\frac{1}{4} 
 {\cal F}^{'I}_{ij}  {\cal F} ^{'I ij }
\\
\nonumber
 -e^{-4\psi/\sqrt{3}+2\sqrt{2/3} \phi}\frac{1}{4} D_i\Phi^I D^i\Phi^I 
+\frac{1}{8}e^{-2\sqrt{2/3}\phi-2\psi/\sqrt{3}} 
\Big], 
\end{eqnarray}
(one can see  that (\ref{new-CV-4D-action}) differrs also 
from the bosonic truncation of the ${\cal N}=4,~D=4$ Freedman-Schwarz gauged supergravity model used in \cite{Chamseddine:1997nm,Chamseddine:1998mc}).

The four-dimensional line element reads
\begin{eqnarray}
\label{new-CV1}
ds^2=r_0^3 e^{3 \nu}\sqrt{\Lambda}\left(-dt^2+dr^2+Y(d\theta^2
+\frac{\sin^2 \theta d \varphi^2}{\Lambda})\right),~~{\rm with~~}
\Lambda= 1+B_0^2Y \sin^2 \theta,
\end{eqnarray}
while the expression of the new dilaton $\psi$ and the nonvanishing U(1) potential is
\begin{eqnarray}
\label{new-CV2}
e^{a\psi}=r_0 e^{\nu}\sqrt{\Lambda},~~
{\cal W}_{i}dx^i=\frac{B_0Y \sin^2 \theta }{2\Lambda}d\varphi.
\end{eqnarray}
The four-dimensional YM and Higgs fields are still given by (\ref{A4}), (\ref{higgs}),
with $w=r/\sinh r$.

One can easily see that for $B_0=0$ the Chamseddine-Volkov
solution is recovered, since the scalars $\phi,\psi$ are not independent 
in this case.
Asymptotically, the geometry (\ref{new-CV1}) approaches the
Melvin-type solution in  ${\cal N}=4,~D=4$ gauged supergravity  found in \cite{Radu:2003av}.
Therefore we interpret the solution (\ref{new-CV1})-(\ref{new-CV2}) as describing 
a nonabelian soliton in a magnetic universe.
The same procedure can be applied to the more 
general globally regular and black hole solutions in \cite{Gubser:2001eg}. 

\section{Further remarks}

The main purpose of this paper was to propose a generalization of the known $D=4$ spherically symmetric  
nonabelian solutions by including the effects of a background U(1) magnetic field.
In this case, the resulting configurations have axial symmetry 
and  approach asymptotically a dilatonic Melvin background.
In our approach, we have used a twisting procedure applied to a set of five-dimensional
configurations in EYM theory.
It would be interesting to construct this type of solutions for a simpler version of the action
than (\ref{action4}), 
without making use of the twisting procedure;
 however, this would require to solve a complicated set of partial differential equations with suitable
 boundary conditions.

More complicated solutions with Melvin-type asymptotics
  in EYM-Higgs-U(1)-dilaton theory are found by starting with other static EYM  
configurations instead of (\ref{BK}), (\ref{YMans}).
The general procedure works as follows:
one starts with an axially symmetric EYMd ($a=\sqrt{3}$) solution $(\gamma^{0}_{ij},~A_i^{(0)I},~\psi^{0})$,
where $\gamma^{0}_{ij}dx^i dx^j=d \ell^2+\gamma_{\varphi \varphi}^0d\varphi^2$,
and uplifts it to $D=5$ according to (\ref{KK5D}).
After twisting and reducing back to four dimensions, one generates in this way 
a new configuration with
\begin{eqnarray}
\label{n1}
ds^2=\gamma_{ij}dx^i dx^j=\sqrt{\Lambda}(d\ell^2+\frac{ \gamma_{\varphi \varphi}^0}{\Lambda }d\varphi^2),
 ~~{\rm with~~} \Lambda=1+ e^{-3a\psi_0}B_0^2\gamma_{\varphi \varphi}^0,~~e^{2a\psi}=e^{2a\psi_0}\Lambda ,
\\
\nonumber
{\cal W}_{i}dx^i=\frac{e^{-3 a\psi} B_0}{2\Lambda }\gamma_{\varphi \varphi}^0 d{\varphi},~~~
\Phi^I=B_0 A_\varphi^{(0)I},~~~A_i^Idx^i=A_i^{(0)I}dx^i-2B_0  A_{\varphi}^{(0)I} {\cal W}_{i}dx^i.
\end{eqnarray}
For example, the $D=4$ EYM(-dilaton) theory possesses also static axially symmetric black hole solutions
\cite{Kleihaus:1997ws,Ibadov:2005rb}, with 
(these configurations are not known in closed form)
\begin{eqnarray}
\label{n2}
d\ell^2= 
  - f dt^2 +  \frac{m}{f} d r^2 + \frac{m r^2}{f} d \theta^2,~~ 
\gamma_{\varphi \varphi}^0 = \frac{l r^2 \sin^2 \theta}{f}, 
\end{eqnarray}
where the metric functions
$f$, $m$ and $l$ are functions of 
the coordinates $r$ and $\theta$, only.
After a suitable gauge transformation, the SU(2) matter fields of these solutions are written in terms of four 
potentials $H_i(r,\theta)$ as  
\begin{eqnarray}
\label{n3}
 A_i^{(0)} dx^i=  
  n \sin \theta \left( H_3 \tau_3 +(1-H_4) \tau_1 \right)   d \varphi
- \left((H_1/r) dr + (1-H_2) d\theta \right) \tau_2
 +\tau_2 d\theta + n\tau_3 \cos \theta d \varphi - n \tau_1 \sin \theta d \varphi.
 \end{eqnarray}
These asymptotically flat solutions are characterized  by their horizon radius and  
three positive integers $(k,n,p)$, where $k$ is related to the polar angle, $n$ to the azimuthal angle   
and $p$ to the node number of some gauge functions
(the spherically symmetric solutions have $k=n=1$). 
 As $r_h\to 0$, a nontrivial globally regular solution is approached \cite{Ibadov:2004rt}.
By using this type of seed solutions one can construct more general 
axially symmetric configurations
describing vortex ring solutions in a background U(1)
magnetic field, where these vortex ring solutions need not only be located
in the $xy$-plane, but might also come in pairs located symmetrically
above and below the $xy$-plane (similar to the asymptotically
flat vortex ring solutions \cite{Kleihaus:2003xz}).
Similar to the asymptotically flat case, one expects all these configurations to be unstable.

Fluxbrane solutions with with nonabelian fields in  
$4+N$ spacetime dimensions can be generated in a similar way, by starting again with solutions of the Eqs. (\ref{eqs}) (the dilaton coupling constant there would depend on $N$).
Also, a similar construction to that presented in this paper can be done starting with a more complicated
higher dimensional action instead of (\ref{action5}),
a particularly interesting case being the $D=10$ low energy heterotic string theory action,
which contains nonabelian fields in the bulk.

The Einstein-Maxwell-dilaton theory has also a solution
describing a pair of oppositely charged
black holes in an external gauge field \cite{Ernst2,Dowker:1993bt}. 
Its euclideanised version describes
the analogue of the Schwinger pair production
of charged particles in a uniform electromagnetic field \cite{Dowker:1993bt}.
It would be interesting to construct the nonabelian counterparts of these configurations.
\\
\\
{\bf Acknowledgement}
\\
BK gratefully acknowledges support by the German Aerospace Center.
The work of ER was supported by a fellowship from the Alexander von Humboldt Foundation. 


\end{document}